\documentclass[10pt,DIV9,a4paper,final]{scrartcl}

\newif\ifartclversion\artclversiontrue
\newif\iflncsversion
\newif\ifjournalversion

\usepackage{amsmath}
\usepackage{amssymb}

\usepackage[english]{babel}

\usepackage[latin1]{inputenc}
\usepackage[T1]{fontenc}

\usepackage[final]{microtype}
\usepackage{enumitem}
\usepackage{ellipsis}
\usepackage{shuffle}
\usepackage{ifthen}
\usepackage{xspace}
\usepackage{xcolor}
\usepackage{url}

\usepackage{booktabs}
\usepackage{multirow}

\usepackage{tikz}

\usetikzlibrary{automata}
\tikzstyle{state without output}=  [circle,draw,minimum size=1.5em,every state,inner sep=0pt]
\tikzstyle{state with output}=     [circle split,draw,minimum size=1.5em,every state,inner sep=0pt]
\tikzset{node distance=1.7cm}

\usepackage[final]{hyperref}

\newenvironment{mk}{\noindent\color{blue} MK:} {}
\newcommand{\mkk}[1]{\begin{mk} #1\end{mk}\xspace}

\newenvironment{al}{\noindent\color{red} AL:} {}
\newcommand{\alx}[1]{\begin{al} #1\end{al}\xspace}

\newenvironment{fj}{\noindent\color{orange} FJ:} {}

\newcommand{\set}[2]{\left\{#1\mathrel{\left|\vphantom{#1}\vphantom{#2}\right.}#2\right\}}
\newcommand{\oneset}[1]{\left\{\mathinner{#1}\right\}}
\newcommand{\smallset}[1]{\left\{#1\right\}}

\newcommand{\union}{\mathbin{\cup}}
\newcommand{\bigunion}{\mathop{\bigcup}}
\newcommand{\isect}{\mathbin{\cap}}

\newcommand{\dotcup}{\ensuremath{\dot\cup}}
\newcommand{\bigdotcup}{\ensuremath{\dot\bigcup}}

\newcommand{\disjunion}{\mathbin{\dotcup}}
\newcommand{\bigdisjunion}{\mathop{\bigdotcup}}

\let\iff=\undefined
\let\implies=\undefined
\let\impliedby=\undefined
\newcommand{\iff}       {\mathrel{\Leftrightarrow}}
\newcommand{\implies}   {\mathrel{\Rightarrow}}

\newcommand{\impliedby} {\mathrel{\Leftarrow}}

\newcommand{\nin}       {\not\in}

\newcommand{\abs}[1]{\left|\mathinner{#1}\right|}

\newcommand{\N}{\mathbb{N}}

\newcommand{\B}{\mathbb{B}}

\newcommand{\calA}{\mathcal{A}}
\newcommand{\calB}{\mathcal{B}}
\newcommand{\calC}{\mathcal{C}}

\newcommand{\calT}{\mathcal{T}}
\newcommand{\calI}{\mathcal{I}}

\newcommand{\calDA}{\mathcal{D\hspace{-0.35mm}A}}

\newcommand{\logicfont}[1]{\mathrm{#1}}

\newcommand{\FO}{\logicfont{FO}}

\newcommand{\TL}{\logicfont{TL}}  

\newcommand{\X}{\mathsf{X}}
\newcommand{\Y}{\mathsf{Y}}

\newcommand{\limplies}   {\mathbin{\rightarrow}}

\newcommand{\liff}       {\mathbin{\leftrightarrow}}

\renewcommand{\land}{\mathbin{\wedge}}

\newcommand{\Synt}{\mathrm{Synt}}

\newcommand{\DA}{\mathbf{D\hspace*{-0.25mm}A}}

\newcommand{\greenfont}[1] {\ensuremath{\mathcal{#1}}}

\newcommand{\greenR} {\greenfont{R}\xspace}
\newcommand{\Requiv} {\mathrel{\greenR}}
\newcommand{\Req}    {\Requiv}
\newcommand{\Rleq}   {\mathrel{\leq_\greenR}}
\newcommand{\Rl}     {\mathrel{<_\greenR}}

\newcommand{\greenL} {\greenfont{L}\xspace}
\newcommand{\Lequiv} {\mathrel{\greenL}}
\newcommand{\Leq}    {\Lequiv}
\newcommand{\Lleq}   {\mathrel{\leq_\greenL}}

\newcommand{\greenJ} {\greenfont{J}\xspace}
\newcommand{\Jequiv} {\mathrel{\greenJ}}
\newcommand{\Jeq}    {\Jequiv}
\newcommand{\Jleq}   {\mathrel{\leq_\greenJ}}

\let\alph=\undefined
\newcommand{\alph}{\alp}
\newcommand{\alp}{\mathrm{alph}}

\newcommand{\sqleq}{\sqsubseteq}

\ifartclversion
\newtheorem{theorem}{Theorem}

\newtheorem{proposition}[theorem]{Proposition}
\newtheorem{lemma}[theorem]{Lemma}
\newtheorem{corollary}[theorem]{Corollary}

\newtheorem{remark}[theorem]{Remark}
\newenvironment{proof}[1][]{\pagebreak[3]\noindent\textit{Proof\ifthenelse{\equal{#1}{}}{}{ (#1)}. }}{\pagebreak[3]\medskip}
\newtheorem{expl}{Example}

\newcommand{\qed}{\hspace*{\fill}\ensuremath{\Box}}
\def\squareforqed{\hbox{\scalebox{.88}{$\Box$}}}
\newcommand{\eex}{\hspace*{\fill}\rotatebox{45}{\scalebox{.71}{\squareforqed}}}
\newenvironment{appresult}[1]{%
  \renewcommand{\label}[1]{}
  \smallskip
  \noindent%
  \textbf{{#1}.} \it}{}
\else
\def\squareforqed{\hbox{\scalebox{.88}{$\Box$}}}
\newcommand{\eex}{\hspace*{\fill}\rotatebox{45}{\scalebox{.71}{\squareforqed}}}

\fi

\setlist{itemsep=0pt,parsep=0pt,topsep=3pt}

\newcommand{\leow}{\triangleright}
\newcommand{\reow}{\triangleleft}

\newcommand{\staigerAcceptance}{\calT}

\newcommand{\potwo}{\text{po2}\xspace}

\newcommand{\trans}[1]{\mathrel{%
    \tikz[baseline=-0.35ex]\draw[->,line cap=round] (0,0) -- %
    node[above,inner sep=0pt,outer sep=2pt]{\ensuremath{\scriptstyle
        #1}} %
    (3.5mm, 0); %
  }}

\newcommand{\longtrans}[1]{\mathrel{%
    \tikz[baseline=-0.35ex]\draw[->,line cap=round] (0,0) -- %
    node[above,inner sep=0pt,outer sep=2pt]{\ensuremath{\scriptstyle
        #1}} %
    (5.5mm, 0); %
  }}

\newcommand{\vdsh}{\mathrel{\makeatletter%
  \def\vd@sh{\raisebox{-.15ex}{$\vdash$}}%
  \mathchoice{\vd@sh}{\vd@sh}%
  {\text{\scriptsize\ensuremath{\vd@sh}}}%
  {\text{\tiny\ensuremath{\vd@sh}}}%
  \makeatother}}

\def\clap#1{\hbox to 0pt{\hss#1\hss}}

\newcommand{\abstractcontent}{We consider ideals and Boolean
  combinations of ideals. For the regular languages within these
  classes we give expressively complete automaton models. In addition,
  we consider general properties of regular ideals and their
  Boolean combinations. These properties include effective algebraic
  characterizations and lattice identities.

  In the main part of this paper we consider the following
  deterministic one-way automaton models: unions of flip automata,
  weak automata, and Staiger-Wagner automata. We show that each of
  these models is expressively complete for regular Boolean
  combination of right ideals. Right ideals over finite words resemble
  the open sets in the Cantor topology over infinite words.  An
  omega-regular language is a Boolean combination of open sets if and
  only if it is recognizable by a deterministic Staiger-Wagner
  automaton; and our result can be seen as a finitary version of this
  classical theorem. In addition, we also consider the canonical
  automaton models for right ideals, prefix-closed languages, and
  factorial languages.

  In the last section, we consider a two-way automaton model which is
  known to be expressively complete for two-variable first-order
  logic.  We show that the above concepts can be adapted to these
  two-way automata such that the resulting languages are the right
  ideals (resp.\ prefix-closed languages, resp.\ Boolean combinations
  of right ideals) definable in two-variable first-order logic.}

\ifartclversion
\begin{document}

\title{Regular Ideal Languages %
  \texorpdfstring{\\}{}%
  and Their Boolean Combinations}%
\author{Franz Jahn \and %
  Manfred Kuf\-leitner%
  \thanks{The last two authors were supported by the German Research
    Foundation (DFG) under grant \mbox{DI 435/5-1}.}%
  \textsuperscript{\phantom{\fnsymbol{footnote}}} %
  \and Alexander Lauser\textsuperscript{\fnsymbol{footnote}}}%
\date{\large FMI, University of Stuttgart, Germany\\[3pt]
  \normalsize
  \texttt{jahnfz@studi.informatik.uni-stuttgart.de} \\
  \texttt{\{kufleitner,\,lauser\}@fmi.uni-stuttgart.de}}%
\maketitle

\begin{abstract}
  \noindent \textbf{Abstract. }  
  \abstractcontent
  
\end{abstract}
\else

\begin{document}

\title{Regular Ideal Languages%
  \texorpdfstring{\\}{}%
  and Their Boolean Combinations}%
\titlerunning{Regular Ideal Languages and Their Boolean Combinations}

%
%
\author{Franz Jahn \and%
  Manfred Kuf\-leitner%
  \thanks{The last two authors were supported by the German Research
    Foundation (DFG) under grant \mbox{DI
      435/5-1}.}
  \and Alexander Lauser\textsuperscript{\fnsymbol{footnote}}%
}%

\authorrunning{F.~Jahn \and M.~Kuf\-leitner \and A.~Lauser}

\institute{FMI, University of Stuttgart, Germany \\
  \footnotesize \texttt{jahnfz@studi.informatik.uni-stuttgart.de} \\
  \footnotesize \texttt{\{kufleitner,\,lauser\}@fmi.uni-stuttgart.de}}

%
%


\maketitle

\begin{abstract}
  \abstractcontent
%
\end{abstract}

\fi

\section{Introduction}

The Cantor topology over infinite words is an important concept for
classifying languages over infinite words. For example, an
$\omega$-regular language is deterministic if and only if it is a
countable intersection of open sets,
\textit{cf.}~\cite[Remark~5.1]{tho90handbook:short}. There are many
other properties of $\omega$-languages which can be described using
the Cantor topology, see
\textit{e.g.}~\cite{pp04:short,sta97handbook:short}.  Ideals are the
finitary version of open sets in the Cantor topology.  A subset $P$ of
a monoid $M$ is a right (resp.\ left, two-sided) \emph{ideal} if $PM
\subseteq P$ (resp.\ $MP \subseteq P$, $MPM \subseteq P$). In
particular, a language $L \subseteq A^*$ is a right ideal if $L A^*
\subseteq L$. A \emph{filter} is the complement of an ideal. Thus over
finite words, a language $L \subseteq A^*$ is a right filter if and
only if it is \emph{prefix-closed}, \textit{i.e.}, if $uv \in L$
implies $u \in L$. Prefix-closed languages correspond to closed sets
in the Cantor topology. A language $L \subseteq A^*$ is a two-sided
filter if and only if it is \emph{factorial} (also known as
\emph{factor-closed} or \emph{infix-closed}), \textit{i.e.}, if $uvw
\in L$ implies $v \in L$.  Our first series of results gives effective
algebraic characterizations of right (resp.\ left, two-sided) ideal
languages and of Boolean combinations of such languages. In addition,
we give lattice identities for each of the resulting language classes.
As a byproduct, we show that a language is both regular and a Boolean
combination of right (resp.\ left, two-sided) ideals if and only if it
is a Boolean combination of regular right (resp.\ left, two-sided)
ideals, \textit{i.e.}, if $\calI$ is the class of right (resp.\ left,
two-sided) ideals and $\mathrm{REG}$ is the class of regular
languages, then $\mathrm{REG} \cap \B\calI = \B(\mathrm{REG} \cap
\calI)$. Here, $\B$ denotes the Boolean closure.

The second contribution of this paper consists of expressively
complete (one-way) automaton models for right ideals, for
prefix-closed languages, for factorial languages, and for Boolean
combinations of right ideals. The results concerning ideals and closed
languages are straightforward and stated here only to draw a more
complete picture. Our main original contribution are automaton models
for regular Boolean combinations of right ideals.  We always assume
that every state in an automaton is reachable from some initial state,
\textit{i.e.}, all automata in this paper are accessible.
\begin{itemize}
\item A \emph{flip automaton} is an automaton with no transitions from
  final states to non-final states, \textit{i.e.}, it ``flips'' at
  most once from a non-final to a final state.  Consequently, every
  minimal complete flip automaton has at most one final state which
  has a self-loop for each letter of the alphabet.  Paz and Peleg have
  shown that if a language~$L$ is recognized by a complete
  deterministic automaton $\calA$, then~$L$ is a right ideal if and
  only if $\calA$ is a flip automaton~\cite{PazPeleg65jacm:short}.  A
  language is a regular Boolean combination of right ideals if and
  only if it is recognized by a union of flip automata (which do not
  have to be complete).
\item An automaton is \emph{fully accepting} if all states are
  final. A word $u$ is rejected in a fully accepting automaton~$\calA$
  if and only if there is no $u$-labeled path in~$\calA$ which starts
  in an initial state. Nondeterministic fully accepting automata are
  expressively complete for prefix-closed languages.  Moreover, if a
  language~$L$ is recognized by a deterministic trim automaton
  $\calA$, then~$L$ is prefix-closed if and only if $\calA$ is fully
  accepting.
\item A \emph{path automaton} is an automaton $\calA$ such that all
  states are both initial and final, \textit{i.e.}, a word $u$ is
  accepted by $\calA$ if there exists a $u$-labeled path
  in~$\calA$. Both deterministic and nondeterministic path automata
  recognize exactly the class of regular factorial languages. This
  characterization can be implicitly found in the work of
  Avgustinovich and Frid~\cite{AvgustinovichFrid06csr:short}.
\item An automaton is \emph{weak} if in each strongly connected
  component either all states are final or all states are
  non-final. Any run of a weak automaton flips only a bounded number
  of times between final and non-final states. Nondeterministic weak
  automata can recognize all regular languages. On the other hand, if
  a language $L$ is recognized by a deterministic automaton~$\calA$,
  then $L$ is a Boolean combination of right ideals if and only
  if~$\calA$ is weak.  Weak automata have been introduced by Muller,
  Saoudi, and Schupp~\cite{mss86:short}.
\item \emph{Deterministic Staiger-Wagner automata} over infinite words
  have been used for characterizing $\omega$-languages $L \subseteq
  A^\omega$ such that both $L$ and $A^\omega \setminus L$ are
  deterministic~\cite{sw74eik:short}. Acceptance of a run in a
  Staiger-Wagner automaton only depends on the set of states visited
  by the run (but not on their order or their number of
  occurrences). We show that, over finite words, deterministic
  Staiger-Wagner automata are expressively complete for Boolean
  combinations of right ideals.  In particular, deterministic
  Staiger-Wagner automata and deterministic weak automata accept the
  same class of languages.
\end{itemize}
We note that flip automata, fully accepting automata, and weak
automata yield effective characterizations of the respective language
classes. For example, in order to check whether a deterministic
automaton $\calA$ recognizes a Boolean combination of right ideals, it
suffices to test if $\calA$ is weak. Moreover, the above automaton
models can easily be applied to subclasses of automata such as
counter-free automata~\cite{mp71:short}. This immediately yields
results of the following kind: A regular language $L$ is both
star-free and a Boolean combination of right ideals if and only if its
minimal automaton is weak and counter-free.

For some classes of languages it is more adequate to use two-way
automata. The relation between two-way automata and ideals (resp.\
closed languages, Boolean combinations of ideals) is more complex than
for one-way automata.  In the last section, we consider deterministic
partially ordered two-way automata (po2dfa). Partially ordered
automata are also known as \emph{very weak}, \emph{1-weak}, or
\emph{linear} automata.  We give restrictions of po2dfa's which define
the right ideals (resp.\ prefix-closed languages, Boolean combinations
of right ideals) inside the po2dfa-recognizable languages.  The class
of languages recognized by po2dfa has a huge number of equivalent
characterizations; these include the variety $\DA$ of finite monoids,
two-variable first-order logic, unary temporal logic, unambiguous
polynomials, and rankers; see
\textit{e.g.}~\cite{tt02:short,dgk08ijfcs:short}. Some of these
characterizations admit natural restrictions which are expressively
complete for their ideal (resp.\ prefix-closed, Boolean combination of
ideals) counterparts. We introduce one-pass flip po2dfa (resp.\
one-pass fully accepting po2dfa, one-pass po2dfa) as expressively
complete automaton models for right ideals (resp.\ prefix-closed
languages, Boolean combinations of right ideals) inside the class of
po2dfa-recognizable languages.  For definitions of these automaton
models, we refer the reader to Section~\ref{sec:DA}. The main
challenge for each of the above automaton models is showing closure
under union and intersection since standard techniques, such as
sequentially executing one automaton after the other, cannot be
applied.
As a complementary result we see that weak one-pass two-way dfa's have
the same expressive power as their one-way counterparts,
\textit{i.e.}, recognize regular Boolean combinations of right ideals.

\section{Preliminaries}

Throughout this paper, $A$ is a finite alphabet. The set of finite
words over the alphabet~$A$ is denoted by $A^*$; it is the free monoid
over $A$.  The neutral element is the empty word~$\varepsilon$. The
set of nonempty words is $A^+ = A^* \setminus
\smallset{\varepsilon}$. If a language $L
\subseteq A^*$ satisfies $L A^* \subseteq L$ (resp.\ $A^* L \subseteq L$,
$A^* L A^* \subseteq L$), then $L$ is a \emph{right
  ideal} (resp.\ \emph{left ideal}, \emph{two-sided ideal}).
If $L = A^* \setminus K$ for some right (resp.\ left, two-sided) ideal
$K$, then $L$ is \emph{prefix-closed} (resp.\ \emph{suffix-closed},
\emph{factorial}). Factorial languages are also known as
\emph{factor-closed} or \emph{infix-closed}.
Boolean combinations consist of complementation,
\emph{finite} unions, and \emph{finite} intersections. 

\emph{Green's relations} on a monoid $M$ are defined as follows.  For
$x, y \in M$ let $x \Rleq y$ (resp.\ $x \Lleq y$, $x \Jleq y$)
if there exist $s,t \in M$ such that $x = ys$ (resp.\ $x = ty$, $x = tys$). We set $x \Req y$ if both $x \Rleq y$ and $y \Rleq
x$. The relations $\Leq$ and $\Jeq$ are defined similarly involving
$\Lleq$ and $\Jleq$, respectively.
An element $x \in M$ is \emph{idempotent} if $x = x^2$. In every
finite monoid~$M$ there exists a number $\omega \geq 1$ such that
$x^\omega$ is idempotent for all $x \in M$.
A homomorphism $h : A^* \to M$ \emph{recognizes} a language~$L
\subseteq A^*$ if~$L = h^{-1}(P)$ for some $P \subseteq M$, \textit{i.e.}, $u
\in L$ if and only if $h(u) \in P$. A monoid~$M$ recognizes~$L$ if
there exists a homomorphism $h: A^* \to M$ recognizing~$L$. For every
regular language $L$ there exists a unique minimal finite monoid
$\Synt(L)$ which recognizes $L$ (and which is effectively computable
as the transition monoid of the minimal automaton). It is the
\emph{syntactic monoid} of $L$, and it is naturally equipped with a
recognizing homomorphism $h_L : A^* \to \Synt(L)$, called the
\emph{syntactic homomorphism}. A language is regular if and only if
its syntactic monoid is finite, see \textit{e.g.}~\cite{pp04:short}.

Lattice identities are a tool for describing classes of languages
(these language classes form so-called lattices). Lattice identities
can be defined in the general setting of free profinite
monoids~\cite{ggp08icalp:short}. In this paper, we only introduce the
$\omega$-notation. We inductively define $\omega$-terms over a set of
variables~$\Sigma$: Every $x \in \Sigma$ is an $\omega$-term; and if
$x$ and $y$ are $\omega$-terms, then so are $xy$ and $(x)^\omega$.
For a number $n \in \N$ and an $\omega$-term $x$, we define $x(n)$
inductively by $(xy)(n) = x(n) y(n)$, $(x^{\omega})(n) = x(n)^{n!}$,
and $x(n) = x$ for $x \in \Sigma$, \textit{i.e.}, $x(n)$ is the word
obtained by replacing all exponents $\omega$ in $x$ by $n!$.
Intuitively, $x^\omega$ is the idempotent element generated by $x$
with respect to \emph{all} regular languages.
A regular language~$L$ satisfies the lattice identity $x \limplies y$
for $\omega$-terms $x$ and $y$ if there exists $n_0 \in \N$ such that
for all $n \geq n_0$ 
and for all homomorphisms $h : \Sigma^* \to A^*$ the implication
$h\big(x(n)\big) \in L \,\implies\, h\big(y(n)\big) \in L$ holds. It
satisfies $x \liff y$ if 
$x \limplies y$ and $y \limplies x$.


\section{Ideals and Their Boolean Combinations}
\label{sec:GandBG}

Many interesting properties over finite words can be stated as
follows: There exists a prefix (resp.\ suffix, factor) which has some
desirable property~$L \subseteq A^*$ and we do not care about
subsequent actions. This immediately leads to the right ideal $LA^*$
(resp.\ left ideal $A^* L$, two-sided ideal $A^* L A^*$). Such
languages and their Boolean combinations arise naturally, see
\textit{e.g.}~\cite{bp91tcs:short,kl11afl:short}. We give effective
algebraic characterizations and lattice identities for the regular
ideal languages (Proposition~\ref{prp:cantor}) and the regular Boolean
combinations of ideals (Theorem~\ref{thm:boolcantor}). In the case of
ideals, the proof is straightforward and relies on the following
simple fact. If $h : M \to N$ is a surjective homomorphism between
monoids and $I \subseteq M$ as well as $J \subseteq N$ are right
ideals (resp.\ left ideals, two-sided ideals), then $h(I)$ and
$h^{-1}(J)$ are also right ideals (resp.\ left ideals, two-sided
ideals), \textit{i.e.}, ideals are closed under homomorphic and
inverse homomorphic images.

\begin{proposition}\label{prp:cantor}
  Let $L \subseteq A^*$ be a regular language recognized by a
  surjective homomorphism $h: A^* \to M$ onto a monoid $M$. The
  following are equivalent:
  \begin{enumerate}
  \item\label{aaa:cantor} %
    $L$ is a right ideal (resp.\ left ideal, two-sided ideal).
  \item\label{ccc:cantor} %
    $h(L)$ is a right ideal (resp.\ left ideal, two-sided ideal).
  \item\label{eee:cantor} %
    $L$ satisfies the lattice identity $y \limplies yz$ (resp.\ $y \limplies
    xy$, $y \limplies xyz$).
  \end{enumerate}
\end{proposition}
\begin{proof} 
  Right ideals (resp.\ left ideals, two-sided ideals) are
  closed under surjective homomorphisms and under inverse
  homomorphisms. Thus~\eqref{aaa:cantor} and~\eqref{ccc:cantor} are
  equivalent.
  We have $L A^* \subseteq L$ if and only if for all $y,z \in A^*$ we
  have $y \in L \implies yz \in L$ if and only if $L$ satisfies the
  lattice identity $y \limplies yz$. This establishes the equivalence
  of~\eqref{aaa:cantor} and~\eqref{eee:cantor} for right ideals; the
  argument for left ideals and two-sided ideals is analogous.
  \qed
\end{proof}

In particular, property~\eqref{ccc:cantor} of
Proposition~\ref{prp:cantor} yields decidability of whether a given
regular language is a (right, left, or two-sided) ideal of $A^*$
because the syntactic homomorphism $h_L : A^* \to \Synt(L)$ and the
set $h_L(L)$ are effectively computable. Moreover, regular (right,
left, and two-sided) ideals are closed under union, intersection, and
inverse homomorphisms. They do not form so-called \emph{positive
  varieties} because they are not closed under residuals (even though
right ideals are closed under left residuals, and left ideals are
closed under right residuals), \textit{cf.}~\cite{pp04:short}. An easy
example is $L = a b A^*$ over the alphabet $A = \smallset{a,b}$; we
have $a \in Lb^{-1} = L \cup \smallset{a}$ and $aa \not\in Lb^{-1}$,
showing that $Lb^{-1}$ is not a right ideal.

In the next theorem, we consider Boolean combinations of ideals.
Note that if $h : M \to N$ is a surjective homomorphism and $I,J$ are
ideals of $M$, then in general, we have $h(I \setminus J) \neq h(I)
\setminus h(J)$. Another obstacle for Boolean combinations of ideals
is the following: If $L$ is regular and a Boolean combination of
ideals $K_i$, then the $K_i$ need not 
be regular. As a byproduct
of our characterization in Theorem~\ref{thm:boolcantor}, we see
that in the above situation, one can find regular ideals $K'_i$ such
that $L$ is a Boolean combination of the languages $K'_i$.

\begin{theorem}\label{thm:boolcantor}
  Let $L \subseteq A^*$ be a language recognized by a surjective
  homomorphism $h: A^* \to\nolinebreak M$ onto a finite monoid
  $M$. Then the following are equivalent:
  \begin{enumerate}
  \item\label{bbb:boolcantor} %
    $L$ is a Boolean combination of right (resp.\ left, two-sided)
    ideals.
  \item\label{ddd:boolcantor} %
    $h(L)$ is a union of $\greenR$-classes (resp.\ $\greenL$-classes,
    $\greenJ$-classes).
  \item\label{fff:boolcantor} %
    $L$ satisfies the lattice identity\, $z(xy)^\omega x \,\liff\,
    z(xy)^\omega$\, (resp.\ the identity
    $s(ts)^\omega z \,\liff\,
    (ts)^\omega z$, the identity $s(ts)^\omega z (xy)^\omega x
    \,\liff\, (ts)^\omega z(xy)^\omega$).  
  \end{enumerate}
\end{theorem}
\begin{proof}
  We show \eqref{bbb:boolcantor}\;$\iff$\;\eqref{ddd:boolcantor} and
  \eqref{ddd:boolcantor}\;$\iff$\;\eqref{fff:boolcantor} for right
  ideals. Left ideals and two-sided ideals are similar.
  For words $u,v \in A^*$ we write $u \equiv v$ if $h(u) = h(v)$.
  
  \eqref{bbb:boolcantor}\;$\implies$\;\eqref{ddd:boolcantor}: %
  Let $L$ Boolean combination of right ideals. Then $L =
  \bigunion_{i=1}^n P_i \setminus Q_i$ for right ideals $P_i$ and
  $Q_i$.
  To see this, we first use De Morgan's law in order to move negations
  inwards so that neither any intersection nor any union is
  negated. Then we perform an induction on the resulting positive
  Boolean expression. For right ideals and negations thereof the claim
  is trivially true. For union the induction step is also clear. Let
  now $L = L_1 \isect L_2$ and let $L_1 = \bigunion_i P_i \setminus
  Q_i$ and $L_2 = \bigunion_j P'_j \setminus Q'_j$. Then $L =
  \bigunion_{i,j} (P_i \isect P'_j) \setminus (Q_i \union Q'_j)$ and
  the claim follows since right ideals are closed under union and
  intersection.
%
  Consider $u,v$ such that $h(u) \Req h(v)$ and let $x,y \in A^*$ such
  that $v \equiv ux$ and $u \equiv vy$.  Suppose $h(u) \in h(L)$. Let
  $u_j = u(xy)^j$ and $v_j = u_j x$. Now, $u_j \equiv u$, $v_j \equiv
  v$, and $u_j$ is a prefix of $v_j$ which in turn is a prefix of
  $u_{j+1}$.
  Every $u_j$ is in $L$ and hence for every $j \in \N$ there exists $i
  \in \oneset{1, \ldots, n}$ such that $u_j \in P_i \setminus Q_i$. By
  the pigeonhole principle there exist $j < k$ with $u_j, u_k \in P_i
  \setminus Q_i$ for some $i \in \oneset{1,\ldots,n}$. Then $v_j \in
  P_i A^* \subseteq P_i$. If $v_j \in Q_i$, then $u_k \in Q_i A^*
  \subseteq Q_i$ and $u_k \nin P_i \setminus Q_i$, a
  contradiction. Thus $v_j \nin Q_i$ and $v_j \in P_i \setminus Q_i
  \subseteq L$. Hence, $h(v) = h(v_j) \in h(L)$.  This shows that
  $h(L)$ is a union of $\greenR$-classes.

  \eqref{ddd:boolcantor}\;$\implies$\;\eqref{bbb:boolcantor}: %
  Let $R$ be an $\greenR$-class of $M$. Consider the two right ideals
  $R' = \set{x}{x \Rleq R}$ and $R'' = \set{x}{x \Rl R}$.  Then
  $h^{-1}(R) = h^{-1}(R') \setminus h^{-1}(R'')$ is a Boolean
  combination of right ideals (since right ideals are closed under
  inverse homomorphisms). With $h(L)$ being a finite union of
  $\greenR$-classes, the claim follows.

  \eqref{ddd:boolcantor}\;$\implies$\;\eqref{fff:boolcantor}: %
  Suppose $h(L)$ is a union of $\greenR$-classes.
  For every sufficiently large $n \geq 1$ we have $h\bigl( z (xy)^n
  \bigr) \Req h\bigl(z (xy)^n x \bigr)$ for all $x,y,z \in A^*$. Thus
  $z (xy)^n \in L \iff z (xy)^n x \in L$, showing the lattice
  identity.

  \eqref{fff:boolcantor}\;$\implies$\;\eqref{ddd:boolcantor}: %
  Suppose $h(w) \Req h(z) \in h(L)$. Then there exist $x,y \in A^*$
  such that $z \equiv wy$ and $w \equiv zx$. We have $z \equiv z
  (xy)^n$ for all $n \in \N$. Hence, $z (xy)^n \in L$. Choosing~$n$
  sufficiently large, the lattice identity yields $w \equiv z (xy)^n x
  \in L$ and $h(w) \in h(L)$, showing that $h(L)$ is a union of
  $\greenR$-classes.
  \qed
\end{proof}

Since Theorem~\ref{thm:boolcantor}~\eqref{ddd:boolcantor} can be
verified effectively for the syntactic homomorphism, it is decidable
whether a given regular language is a Boolean combination of right
ideals (resp.\ left ideals, two-sided ideals).

Every $\greenR$-class is the set difference between two right
ideals. Thus if $L$ is a Boolean combination of (arbitrary) right
ideals and if $L$ is recognized by $h : A^* \to M$, then by
Theorem~\ref{thm:boolcantor}, the language $L$ can also be written as
a Boolean combination of right ideals $K_i$ such that each $K_i$ is
recognized by $h$. The situation for Boolean combinations of left
ideals (resp.\ two-sided ideals) is similar.

For finite monoids, $\greenJ$ is the smallest equivalence relation
such that $\greenR \subseteq \greenJ$ and $\greenL \subseteq \greenJ$,
see \textit{e.g.}~\cite[Proposition~A.2.5\,(2)]{pp04:short}.  Hence, it
follows from Theorem~\ref{thm:boolcantor} that a regular language
$L$ is a Boolean combination of two-sided ideals if and only
if $L$ is both a Boolean combination of right ideals and a
Boolean combination of left ideals.
\ifjournalversion\mkk{TODO: Bsp, das zeigt, dass es fuer nicht-regulaere
  Sprachen nicht gilt?}\fi


In Boolean combinations of right ideals, intuitively speaking, what
happens is that the end of words is ``concealed.'' Appending a new
symbol as an end-marker to a language yields a Boolean combination of
right ideals. Specifically, if $L$ is language over $A \setminus
\oneset{a}$, then $La$ is a Boolean combination of right ideals of
$A^*$ because $La = La A^* \setminus La A^+$. In Section~\ref{sec:DA},
we will avoid this ``revealing'' of the end of the word by the right
end marker by considering one-pass automata.

\section{One-way Automaton Models}
\label{sec:automata}

As usual, an \emph{automaton} $\calA = (Q, A, \delta, Q_0, F)$ is
given by a finite set of states~$Q$, an input alphabet~$A$, a
transition relation $\delta \subseteq Q \times A \times Q$, a set of
initial states $Q_0 \subseteq Q$, and a set of final states $F
\subseteq Q$. For transitions $(p, a, q) \in \delta$ we write $p
\trans{a} q$ and we inductively extend the transition relation to
words: $q \trans{\varepsilon} q$ for all $q \in Q$; and $p
\longtrans{au} q$ if there exists some $r \in Q$ such that $p
\trans{a} r \trans{u} q$. A \emph{run} on a word $a_1 \cdots a_n$ with
$a_i \in A$ is a sequence of states $q_0 q_1 \cdots q_n$ such that
$q_0 \in Q_0$ and $q_{i-1} \trans{a_i} q_i$ for all $i$.  We always
assume that all states are accessible, \textit{i.e.}, for every $q \in
Q$ there exist $q_0 \in Q_0$ and $u \in A^*$ such that $q_0 \trans{u}
q$. A word $u \in A^*$ is \emph{accepted} by~$\calA$ if there exist $p
\in Q_0$ and $q \in F$ such that $p \trans{u} q$. The language
\emph{recognized} by~$\calA$ is $L(\calA) = \set{u \in A^*}{u \text{
    is accepted by } \calA}$. The automaton $\calA$ is \emph{complete}
if for every $p \in Q$ and for every $a \in A$ there exists at least
one state $q \in Q$ such that $p \trans{a} q$; it is \emph{trim} if
for every $q \in Q$ there exists $u \in A^*$ and $p \in F$ such that
$q \trans{u} p$; and it is \emph{deterministic} if $\abs{Q_0} = 1$ and
for all $p \in Q$ and all $a \in A$ there is at most one state $q \in
Q$ with $p \trans{a} q$.

In the remainder of the section, we give automaton models for regular
right ideals, prefix-closed languages, factorial languages, and
Boolean combinations of right ideals. The results concerning ideals
and closed languages are straightforward and presented here only for
the sake of completeness.  Our main original contribution is
Theorem~\ref{thm:weakdfa}, where we give three automaton descriptions
of Boolean combinations of ideals: deterministic weak automata,
deterministic Staiger-Wagner automata, and unions of deterministic
flip automata.

A \emph{flip automaton} is an automaton such that $p \in F$ and $p
\trans{a} q$ implies $q \in F$. The idea is that, in every run, flip
automata can ``flip'' at most once from non-accepting to
accepting. Note that the language of a complete flip automata remains
unchanged if we add a self-loop $q \trans{a} q$ for every state $q \in
F$ and every letter $a \in A$.

\begin{proposition}\label{prp:flip}
  Let $L \subseteq A^*$ be recognized by a complete deterministic
  automaton~$\calA$. Then the following are equivalent:
  \begin{enumerate}
  \item\label{aaa:flip} $L$ is a right ideal.
  \item\label{bbb:flip} $\calA$ is a flip automaton.
  \item\label{ccc:flip} $L$ is recognized by some complete
    (nondeterministic) flip automaton.
  \end{enumerate}
\end{proposition}
\begin{proof}
  \eqref{aaa:flip}\;$\implies$\;\eqref{bbb:flip}: Let $\calA = (Q, A,
  \delta, q_0, F)$ and suppose $p \trans{a} q$ for $p \in F$ and $a
  \in A$. Since every state is reachable, there exists a word $u \in
  A^*$ such that $q_0 \trans{u} p$. In particular, $u \in L$. Since
  $L$ is a right ideal, we have $ua \in L$. Now, $q_0 \longtrans{ua}
  q$ yields $q \in F$.
  The implication \eqref{bbb:flip}\;$\implies$\;\eqref{ccc:flip} is
  trivial.

  \eqref{ccc:flip}\;$\implies$\;\eqref{aaa:flip}: Let $\calA' = (Q, A,
  \delta, Q_0, F)$ be a complete flip automaton recognizing~$L$.
  Suppose $u \in L$, and let $a \in A^*$ be arbitrary. Then $q_0
  \trans{u} p$ for $q_0 \in Q_0$ and $p \in F$. In addition, since
  $\calA'$ is complete, we have $p \trans{a} q$. Moreover, $q \in F$
  because $\calA'$ is a flip automaton. This shows $LA \subseteq L$
  and thus $LA^* \subseteq L$.
  \qed
\end{proof}

The equivalence of \eqref{aaa:flip} and \eqref{bbb:flip} in
Proposition~\ref{prp:flip} is due to Paz and
Peleg~\cite{PazPeleg65jacm:short}.  Of course, not every complete
nondeterministic automaton which recognizes a right ideal has to be a
flip automaton. Note that arbitrary (\textit{i.e.}, non-complete and
nondeterministic) flip automata can recognize all regular languages.

      
      

A \emph{fully accepting automaton} is an automaton in which all states
are final, \textit{i.e.}, $F = Q$. The only possibility to reject a
word is a missing outgoing transition at some point of the
computation. Complementing Proposition~\ref{prp:flip} leads to the
following characterization of fully accepting automata.


\begin{corollary}\label{cor:coflip}
  Let $L \subseteq A^*$ be recognized by a deterministic trim
  automaton $\calA$. Then the following are equivalent:
  \begin{enumerate}
  \item\label{aaa:coflip} $L$ is prefix-closed.
  \item\label{bbb:coflip} $\calA$ is fully accepting.
  \item\label{ccc:coflip} $L$ is recognized by some (nondeterministic)
    fully accepting automaton.
  \end{enumerate}
\end{corollary}
\begin{proof}
  \eqref{aaa:coflip}\;$\implies$\;\eqref{bbb:coflip}: Let $\calA = (Q,
  A, \delta, q_0, F)$ and assume $p \in Q \setminus F$. Since $\calA$
  is trim, there exist $q \in F$ and $u,v \in A^*$ such that $q_0
  \trans{u} p \trans{v} q$. Now, $uv \in L$ implies $u \in L$ and $p
  \in F$, a contradiction.
  
  The implication \eqref{bbb:coflip}\;$\implies$\;\eqref{ccc:coflip}
  is trivial.

  \eqref{ccc:coflip}\;$\implies$\;\eqref{aaa:coflip}: Let $\calA' =
  (Q, A, \delta, Q_0, F)$ be a nondeterministic fully accepting
  automaton recognizing $L$. Suppose $u \not\in L$, \textit{i.e.}, for
  every $q_0 \in Q_0$ and $q \in Q$ we have $(q_0,u,q) \not\in
  \delta$. Thus for every $v \in A^*$ and every $q_0 \in Q_0$ and $q
  \in Q$ we have $(q_0,uv,q) \not\in \delta$, \textit{i.e.}, we have
  $uv \not\in L$. This shows that $A^* \setminus L$ is a right ideal.
  \qed
\end{proof}

A \emph{path automaton} is an automaton such that every state is both
initial and final, \textit{i.e.}, $Q_0 = F = Q$. In particular, a path
automaton accepts a word $u \in A^*$ if and only if there exists a
path $p \trans{u} q$ for some $p,q \in Q$.

\begin{corollary}\label{cor:pathnfa}
  Let $L \subseteq A^*$ be a regular language.  Then $L$ is factorial
  if and only if~$L$ is recognized by a path automaton.
\end{corollary}
\begin{proof}
  ``$\implies$'': By Corollary~\ref{cor:coflip} (and since $L$ is
  prefix-closed), the language $L$ is recognized by a deterministic
  fully accepting automaton $\calA = (Q, A, \delta, q_0, Q)$. We show
  $L(Q,A,\delta,Q,Q) \subseteq L$. Suppose $p \trans{v} q$ for some $v
  \in A^*$. Then there exists $u \in A^*$ such that $q_0 \trans{u} p$,
  i.e., $uv \in L$. Since $L$ is suffix-closed, we have $v \in L$.
    
  ``$\impliedby$'': Let $\calA = (Q, A, \delta, Q, Q)$ be a path
  automaton with $L(\calA) = L$.  Then $\calA$ as well as the
  automaton obtained by reversing all edges (and interchanging initial
  and final states -- which in this case has no effect) are fully
  accepting. Thus, by Corollary~\ref{cor:coflip}, both the language
  $L$ and its reversal $L^r = \set{a_1 \cdots a_n \in A^*}{a_i \in A,
    \; a_n \cdots a_1 \in L}$ are prefix-closed. It follows that $L$
  is factorial.
  \qed
\end{proof}

For deterministic transition relations, the statement of
Corollary~\ref{cor:pathnfa} can be found implicitly in the work of
Avgustinovich and Frid~\cite{AvgustinovichFrid06csr:short}.

An automaton is \emph{weak} if for every strongly connected
component~$C \subseteq Q$, we either have $C \subseteq F$ or $C \isect
F = \emptyset$. The concept of weak automata has been introduced by
Muller, Saoudi, and Schupp~\cite{mss86:short} for alternating tree
automata. 
A \emph{Staiger-Wagner automaton} is given by $\calB = (Q, A, \delta,
q_0, \staigerAcceptance)$ where $\staigerAcceptance \subseteq 2^Q$.
Acceptance of a run
by a Staiger-Wagner automaton only depends on the set of states
visited by the run. A run $q_0 q_1 \cdots q_n$ is \emph{accepting} if
$\oneset{q_0, q_1, \ldots, q_n} \in \staigerAcceptance$; and a word is
accepted if it has an accepting run.

\begin{lemma}\label{lem:nfa2sw}
  Let $\calA = (Q, A, \delta, Q_0, F)$ be a weak automaton. Then there
  exists $\staigerAcceptance$ such that the Staiger-Wagner automaton
  $\calB = (Q, A, \delta, Q_0, \staigerAcceptance)$ recognizes
  $L(\calA)$.
\end{lemma}
\begin{proof}
  Let $\operatorname{future}(q)$ denote the set of states which are reachable from
  $q$ and which are not located in the same strongly connected
  component as $q$. We can construct $\staigerAcceptance$ as follows:
  \begin{equation*}
    \staigerAcceptance = \set{T}{\exists q \in F \cap T \colon \,
      T \subseteq Q \setminus \operatorname{future}(q)}.
  \end{equation*}
  Each element of $\staigerAcceptance$ guarantees, that a run ends
  within an accepting strongly connected component of $\calA$. Since
  $\calA$ is weak, we conclude $L(\calA) = L(\calB)$.
  \qed 
\end{proof}

Our next result shows that both deterministic weak automata and
deterministic Staiger-Wagner automata are expressively complete for
Boolean combinations of right ideals. Moreover, if a deterministic
automaton $\calA$ recognizes a Boolean combination of right ideals,
then, by Lemma~\ref{lem:nfa2sw}, the automaton $\calA$ itself can be
equipped with a Staiger-Wagner acceptance condition. A third automaton
model for Boolean combinations of right ideals is given by unions of
(not necessarily complete) deterministic flip automata.  This last
property follows from Theorem~\ref{thm:boolcantor} since the inverse
homomorphic image of every $\greenR$-class of a finite monoid is
recognizable by a flip automaton.

\begin{theorem}\label{thm:weakdfa}
  Let $L \subseteq A^*$ be recognized by a deterministic automaton
  $\calA$. Then the following are equivalent:
  \begin{enumerate}
    \item\label{aaa:weakdfa} $L$ is a Boolean combination of right
      ideals.
    \item\label{ccc:weakdfa} $\calA$ is weak.
    \item\label{ddd:weakdfa} $L$ is recognized by some deterministic
      Staiger-Wagner automaton.
    \item\label{eee:weakdfa} $L$ is a finite disjoint union of
      languages $L(\calB_i)$ such that each $\calB_i$ is a
      deterministic flip automaton.
  \end{enumerate}
\end{theorem}
\begin{proof}
  \eqref{aaa:weakdfa}\;$\implies$\;\eqref{ccc:weakdfa}: Let $\calA =
  (Q, A, \delta, q_0, F)$.  Assume $p \trans{x} q$ and $q \trans{y} p$
  for $p \in F$. Choose $z \in A^*$ such that $q_0 \trans{z} p$. Then
  for all $n \in \N$ we have $z(xy)^n \in L$. By
  Theorem~\ref{thm:boolcantor}, the language $L$ satisfies the lattice
  identity $z(xy)^\omega \liff z(xy)^\omega x$. Therefore, for some
  $n$, we have $z(xy)^n x \in L$. Now, $\delta(q_0,z(xy)^nx) = q$
  implies $q \in F$.

  The implication \eqref{ccc:weakdfa}\;$\implies$\;\eqref{ddd:weakdfa}
  follows by Lemma~\ref{lem:nfa2sw}.

  \eqref{ddd:weakdfa}\;$\implies$\;\eqref{aaa:weakdfa}: Let $\calB =
  (Q, A, \delta, q_0, \staigerAcceptance)$ be a deterministic
  Staiger-Wagner automaton. We show that $L(\calB)$ satisfies the
  lattice identity $z(xy)^\omega \liff z(xy)^\omega x$. Let $x,y,z \in
  A^*$ and let $n \geq \abs{Q}$.  Let $q_0 \trans{z} q_1$ and $q_i
  \longtrans{xy} q_{i+1}$ for $1 \leq i \leq n$.  By choice of $n$
  there exist $1 \leq k < \ell \leq n+1$ such that $q_k = q_\ell$.  It
  follows that, for all $m \in \N$, the runs of the words
  $z(xy)^{\ell-1}(xy)^m$ and $z(xy)^{\ell-1}(xy)^m x$ both visit the same
  states as $z(xy)^{\ell-1}$. In particular, $z(xy)^n \in L$ if and
  only if $z(xy)^n x \in L$ (which proves the lattice identity
  $z(xy)^\omega \liff z(xy)^\omega x$ for $L$).
  
  \eqref{ccc:weakdfa}\;$\implies$\;\eqref{eee:weakdfa}: Let $\calA =
  (Q,A,\delta,q_0,F)$ be weak. For a strongly connected component $C
  \subseteq F$ we define $\calB_C = (Q_C, A, \delta_C, q_0, F \isect
  C)$ as the (not necessarily complete) flip automaton with states
  $Q_C = \set{q \in Q}{C \text{ is reachable from } q}$, and its
  transition function $\delta_C$ is the restriction of $\delta$ to
  states $Q_C$. Then $L(\calA) = \bigdisjunion_C L(\calB_C)$ where the
  union ranges over all strongly connected components $C \subseteq
  F$. Note that this union is indeed disjoint because $\calA$ is
  deterministic.

  \eqref{eee:weakdfa}\;$\implies$\;\eqref{ccc:weakdfa}: Every flip
  automaton is weak. Moreover, languages recognized by weak automata
  are closed under union by the equivalence of \eqref{aaa:weakdfa} and
  \eqref{ccc:weakdfa}.
  \qed
\end{proof}


      

Both nondeterministic weak automata and nondeterministic
Staiger-Wagner automata are expressively complete for the class of all
regular languages as the next lemma shows. In particular, the
nondeterministic variants of weak automata and Staiger-Wagner automata
do \emph{not} characterize Boolean combinations of right ideals.

\begin{lemma}\label{lem:weaknfa}
  Let $L \subseteq A^*$. The following are equivalent:
  \begin{enumerate}
  \item\label{aaa:weaknfa} $L$ is regular.
  \item\label{bbb:weaknfa} $L$ is recognized by a (nondeterministic)
    weak automaton.
  \item\label{ccc:weaknfa} $L$ is recognized by a (nondeterministic)
    Staiger-Wagner automaton.
  \end{enumerate}
\end{lemma}
\begin{proof}
  \eqref{aaa:weaknfa}\;$\implies$\;\eqref{bbb:weaknfa}: Let $L$ be
  recognized by the deterministic automaton $\calA =
  (Q,A,\delta,q_0,F)$ and let $f \not\in Q$ be a new state.  Let
  \begin{equation*}
    \delta' = \delta \union \set{(p,a,f)}{(p,a,q) \in \delta 
      \text{ and } q \in F}.
  \end{equation*}
  We set $Q_0 = \smallset{q_0,f}$ if $q_0 \in F$, and $Q_0 =
  \smallset{q_0}$ otherwise. This way we introduce a single accepting
  state $f$ which can be reached nondeterministically if and only if
  there was a path from the initial state to some final state in
  $\calA$.  Thus $(Q,A,\delta',Q_0,\smallset{f})$ recognizes $L$.

  \eqref{bbb:weaknfa}\;$\implies$\;\eqref{ccc:weaknfa}: follows from
  Lemma~\ref{lem:nfa2sw}.

  \eqref{ccc:weaknfa}\;$\implies$\;\eqref{aaa:weaknfa}: Let $\calB =
  (Q, A, \delta, Q_0, \staigerAcceptance)$ be a nondeterministic
  Staiger-Wagner automaton. We can construct $\calA = (2^{Q} \times Q,
  A, \delta', Q_0', F)$ with $((P,q), a, (P',q')) \in \delta'$ if and
  only if $(q, a, q') \in \delta \land P' = P \union \smallset{q'}$.
  The set of initial states is $Q_0' = \set{(\smallset{q}, q)}{q \in
    Q_0}$, and the set of final states is defined by $(P, q) \in F$ if
  and only if $P \in \staigerAcceptance$. This way, $\calA$ simulates
  $\calB$ along each path and collects the visited states. It accepts,
  if the set of visited states is in $\staigerAcceptance$. Therefore
  $L(\calA) = L(\calB)$.
  \qed
\end{proof}

\begin{remark}\label{rem:autodec}
  Proposition~\ref{prp:flip} (resp.\ Corollary~\ref{cor:coflip},
  Theorem~\ref{thm:weakdfa}) yields another decision procedure for the
  class of regular right ideals (resp.\ prefix-closed languages,
  Boolean combinations of right ideals). In the case of
  Proposition~\ref{prp:flip}, this was first observed by Paz and
  Peleg~\cite{PazPeleg65jacm:short}.  Moreover, the above decidability
  results can often be combined with other automaton models. For
  example, a well-known result of McNaughton and Papert says that a
  language is definable in first-order logic if and only if its
  minimal automaton is counter-free~\cite{mp71:short}. Together with
  Theorem~\ref{thm:weakdfa}, we see that a language $L$ is a
  first-order definable Boolean combination of right ideals if and
  only if the minimal automaton of $L$ is weak and counter-free.
  \eex
\end{remark}

\ifjournalversion\alx{Biideale ($K A^* L$) und deren boolesche
  Kombis?}\fi

\section{Two-way Automaton Models and Languages in
  \texorpdfstring{$\mathbf{\mathcal{D\hspace*{-0.5mm}A}}$}{DA}}
\label{sec:DA}

The results in the previous section can easily be translated into
characterizations of regular left ideals (resp.\ suffix-closed
languages, Boolean combinations of left ideals) by considering
automata which read the input from right to left. Varying the
direction of the head movement naturally leads to two-way
automata. The situation for arbitrary two-way automata is more
involved than for one-way automata; the main reason is that two-way
automata are usually defined using left and right end markers. On the
other hand, if $L \subseteq (A \setminus \smallset{a})^*$, then $La =
LaA^* \setminus LaA^+$. This shows that by adding an explicit end
marker, every language becomes a Boolean combination of right
ideals. To overcome this, we introduce the notion of \emph{one-pass}
two-way automata; these automata stop processing the input as soon as
they read the right end marker. Now, the problem with classes of
one-pass two-way automata is that, in general, they may not be closed
under union and intersection (standard techniques, such as executing
one automaton after the other, cannot be applied). We have no
satisfactory solution for arbitrary two-way automata, but we show that
the concepts of Section~\ref{sec:automata} can be adapted to a
well-known subclass of two-way automata, namely deterministic
partially ordered two-way automata (po2dfa). The class of languages
recognized by po2dfa is a natural subclass of the star-free languages
which has a huge number of different characterizations, see
\textit{e.g.}~\cite{tt02:short,dgk08ijfcs:short}. The most prominent
of these characterizations is definability in two-variable first-order
logic. By a description of algebraic means, it is the language variety
$\calDA$, \textit{i.e.}, the class of regular languages satisfying the
lattice identity $p (xy)^\omega q \liff p (xy)^\omega x (xy)^\omega
q$.  As a byproduct, we show that some of the other characterizations
of po2dfa recognizable languages also admit natural counterparts for
right ideals and their Boolean combinations.

\ifjournalversion\alx{$\DA \mapsto \FO^2$ (?). Maybe we can avoid monoid varieties
  completely. Only if there is a (natural) $\FO^2[{<}]$-fragment for
  (Boolean combinations of) right ideals?}\fi

%
A \emph{two-way automaton} is a tuple $\calA = (Z, A, \delta, X_0,
F)$. The finite set of states $Z = X \disjunion Y$ is partitioned into
\emph{right-moving} states $X$ (for ne$X_{\!\!\phantom{.}}$t) and
\emph{left-moving} states $Y$ (for $Y$\!esterday).  The states in $X_0
\subseteq X$ are initial and states in $F \subseteq Z$ are final. On
input $u \in A^*$, the tape content is $\leow u \reow$ where $\leow$
and $\reow$ are new symbols marking the left and right end of the
tape, respectively. Initially, the head is at the first letter of
$u$. The direction in which the input is processed can be controlled
by~$\calA$. The idea is that \emph{before} a transition is made, the
head movement is performed, and the direction of the movement depends
only on the \emph{destination state} of the transition. The left end
marker $\leow$ must not be overrun.  More formally, the transition
relation satisfies $\delta \subseteq (Z \times A \times Z) \union (Y
\times \oneset{\leow} \times X) \union (X \times \oneset{\reow} \times
Z)$.  As for one-way automata, we write $z \trans{a} z'$ instead of
$(z,a,z') \in \delta$.  
More formally, a \emph{configuration} is a pair $(z, i) \in Z \times
\N$ where $z$ is the current state and $i$ is the current position on
the tape. Suppose position~$i$ is labeled by $a \in A \union
\oneset{\leow, \reow}$. Then a transition $(z, i) \vdsh_{\!\calA} (z',
j)$ between configurations exists if $z \trans{a} z'$ and $j = i + 1$
(for $z' \in X$) or $j = i - 1$ (for $z' \in Y$). A \emph{computation}
of $\calA$ on input $u$ is a sequence
\begin{equation*}
  (z_0, i_0) \,\vdsh_{\!\calA}\, \cdots \,\vdsh_{\!\calA}\, (z_t, i_t)
\end{equation*}
of configurations such that $z_0 \in X_0$, $i_0 = 1$, $i_k \in
\oneset{0, \ldots, \abs{u}+1}$ for $1 \leq k < t$, and $i_t =
\abs{u}+2$. Note that position $0$ is labeled with the left end marker
$\leow$ and the position $\abs{u}+1$ is labeled with the right end
marker~$\reow$. The computation is \emph{accepting} if $z_t \in F$ is
final and the input $u$ is accepted if there exists an accepting
computation for it. Note that by the signature of the transition
relation, the left end marker $\leow$ cannot be trespassed.
One-way automata may be seen as special cases with $Y = \emptyset$.
The language $L(\calA)$ \emph{recognized} by~$\calA$ is $L(\calA) =
\set{u \in A^*}{\text{\ensuremath{\calA} accepts \ensuremath{u}}}$.

A two-way automaton is \emph{deterministic} if $\abs{X_0} = 1$ and for
all $z \in Z$ and all $a \in A \union \oneset{\leow, \reow}$ there
exists at most one $z' \in Z$ with $z \trans{a} z'$. For technical
reasons, we also consider the empty automaton ($Z = \delta = X_0 = F =
\emptyset$) as deterministic.  It is
\emph{complete} if for all $z \in Z$ and all $a$ there exists $z' \in
Z$ with $z \trans{a} z'$ (more precisely, we require the existence of
$z'$ if either $z \in Y$ and $a \in A \union \oneset{\leow}$ or if $z
\in X$ and $a \in A \union \oneset{\reow}$).
A two-way automaton is \emph{one-pass} if $z \trans{\reow} z'$ implies
$z = z'$. The idea is that a two-way automaton has finished ``one
pass'' when it encounters the right end marker $\reow$ for the first
time; hence for a one-pass automaton, the acceptance of a word
is determined by the state when scanning $\reow$ for the first time.
The automaton is \emph{partially ordered} if there exists a partial
ordering $\sqleq$ of the states such that transitions are
non-descending, \textit{i.e.}, if $z \trans{a} z'$, then $z \sqleq
z'$. In other words, once a state is left in a partially ordered
automaton, it is never re-entered. We abbreviate ``deterministic
partially ordered two-way automaton'' by \emph{po2dfa}.
%

Schwentick, Th{\'e}rien, and Vollmer~\cite{stv01dlt:short} have shown
that po2dfa are expressively complete for $\calDA$. The main result of
this section is a characterization of Boolean combinations of right
ideals (resp.\ right ideals, prefix-closed languages) in $\calDA$ in
terms of subclasses of one-pass po2dfa. 

A crucial property of one-pass po2dfa is closure under Boolean
combinations; and to see this, we shall need the following
synchronization lemma. The same lemma was formulated already
in~\cite[Lemma~8]{kl11ijfcs:short} for B\"uchi automata and infinite
words. The \emph{alphabet} of a word $u$ is denoted by~$\alp(u)$. The
$i$th letter of $u$ is $u(i)$.
\begin{lemma}[Synchronization Lemma]\label{lem:autopf}
  Consider a po2dfa~$\calA$ with states $Z =
  X \mathbin{\dot\union} Y$. For every $v = a_1 \cdots a_m \in
  \Gamma^+$ there exists a po2dfa~$\calC$ with states $Z_\calC = Z\times
  \smallset{v} \times \smallset{1, \ldots, m}$ such that, for all $u
  \in \Gamma^*$ having a factorization $u = u_1 a_1 \cdots u_m a_m u'$
  with $a_i \nin \alp(u_i)$, the following simulation property holds:
  If
  \begin{equation*}
    (z_0, i_0) \,\vdsh_\calA\, (z_1, i_1) \,\vdsh_\calA\,
    \cdots \,\vdsh_\calA\, (z_n, i_n)
  \end{equation*}
  is a sequence of transitions of $\calA$ for some $n \geq 1$
  with $i_0 = i_n = \abs{u_1 a_1 \cdots u_m a_m}$ and $i_t <
  i_n$ for all $1 \leq t < n$, then
  \begin{equation*}
    \big((z_1, v, k_1) , i_1\big) \,\vdsh_{\calC}\, 
    \cdots \,\vdsh_{\calC}\, 
    \big((z_n, v, k_n), i_n\big)
  \end{equation*}
  is a sequence of transitions of $\calC$ with $k_1 = k_n = m$ such
  that there exists no $1 \leq t < n$ with $z_t \in X$, $k_t = m$, and
  $u(i_t) = a_m$.
  \qed
\end{lemma}

Intuitively, this means that if a deterministic $\potwo$-automaton
moves left at some point in its computation, then it may recognize the
position on the input \emph{on-the-fly}\,---\,provided that this
happens at a suitable position, \textit{i.e.}, the $a_m$ in the
factorization stipulated in Lemma~\ref{lem:autopf}. In the latter
application, determinism will yield such a factorization and for a
partially ordered automaton the parameter $m$ can be bounded over all
inputs $u \in A^*$.
Note that \cite[Lemma~8]{kl11ijfcs:short} was formulated with B\"uchi
automata on infinite words. However, the acceptance condition does not
influence the statement at all and, since the computations considered
in the lemma take place completely on the finite prefix $u_1 a_1
\cdots u_m a_m$, the behavior of the automata is independent of the
suffix $u'$ which may even be an infinite word.

\begin{lemma}\label{lem:det-po2-BooAlg}
  The class of languages recognizable by one-pass po2dfa is a Boolean
  algebra.
%
%
\end{lemma}
\begin{proof}
  Suppose that $\calA = (X \disjunion Y, A, \delta, x_0, F)$ is a
  one-pass po2dfa. By adjoining a new non-final right-moving sink
  state, we may assume that $\calA$ is complete. Then $\calA' = (X
  \disjunion Y, A, \delta, x_0, X \setminus F)$ recognizes $A^*
  \setminus L(\calA)$. Therefore, one-pass po2dfa are closed under
  complementation. It remains to show closure under union.

  We describe a product automaton construction for the union of two
  automaton which executes both automata in parallel. Of course, there
  is only one head to do this, and the main problem to overcome in
  this construction is when the automata disagree on the head
  movement.  We shall only give a high-level description of the
  construction; details can be implemented similarly to the situation
  for deterministic $\potwo$-B{\"u}chi
  automata~\cite{kl11ijfcs:short}.

  By adding a new sink state as needed, we may assume that both automata
  are complete.  The two automata are simulated in parallel as long as
  they agree on moving to the right. This is called the
  \emph{synchronous mode}. If at least one of the automata changes to
  left-moving, then we start a simulation of one of the left-moving
  automata in the so-called \emph{asynchronous mode} while suspending
  the other automaton.  We refer to the position of the input where
  this divergence happens as the \emph{synchronization point}. In
  asynchronous mode, the active automaton can move in either
  direction. As soon as the synchronization point is reached again and
  both automata agree on moving to the right, we switch back to
  synchronous mode and continue simulating both automata in parallel;
  otherwise we stay in asynchronous mode while simulating one of the
  automata.  To implement this idea, the synchronization point must be
  recognized while in the asynchronous mode.
  
  For this re-synchronization, we use Lemma~\ref{lem:autopf} and some
  combinatorial property of computations of po2dfa. Assume that we are
  about to enter the asynchronous mode. Suppose the input~$u$ is
  factorized $u = u_1 a_1 \cdots u_m a_m u'$ such that the $a_i$'s
  correspond to the positions where during synchronous mode at least
  one of the automata changed its state. Note that $a_m$ corresponds
  to the synchronization point because a change from right-moving to
  left moving implies a change of state.  By determinism, we have $a_i
  \nin \alph(u_i)$. Moreover, since both automata are partially
  ordered,~$m$ is bounded by the sum of the number of states of both
  automata. The last observation allows to store the word $v = a_1
  \cdots a_m$ in a bounded stack of letters in the state space. Using
  the automaton from Lemma~\ref{lem:autopf} for $v$ as the active
  automaton, we can simulate the active automaton whilst being aware
  of whenever the synchronization point is reached again. Both
  automata are complete and thus the synchronization point is
  eventually reached by the active automaton. After this, we switch
  back to synchronous mode to simulate both automata in parallel. In
  synchronous mode the stack of letters is administered,
  \textit{i.e.}, whenever a state change happens in one of the
  automata whilst in synchronous mode, the currently scanned letter is
  pushed to the stack. At the end, we accept if one of the automata
  accepts.

  The procedure given above can be done effectively in such a way that
  the simulating automaton is a complete, deterministic one-pass
  \potwo-automata.  The actual construction is along the lines of the
  proof of \cite[Proposition~9]{kl11ijfcs:short} and therefore not
  given here.
  \qed
\end{proof}

A \emph{monomial} is a language 
$P = A_1^* a_1 \cdots A_k^* a_k A_{k+1}^*$ where $A_i \subseteq A$ and
$a_i \in A$. It is \emph{unambiguous} if every word $u \in P$ has a
unique factorization $u = u_1 a_1 \cdots u_k a_k u_{k+1}$ with $u_i
\in A_i^*$.
A convenient intermediate step from languages in $\calDA$ to automata
are rankers. A \emph{ranker} is a word in $\set{\X_a, \Y_{\!a}}{a \in
  A}^*$. Intuitively, a ranker~$r$ represents a sequence of
instructions $\X_a$ for ``
next $a$-position'' and $\Y_{\!a}$ for ``
previous $a$-position'' which is processed from left to right. That
is, for a word $u = a_1 \cdots a_n$ with $a_j \in A$ and a position $i
\in \oneset{0,\ldots,n+1}$ we set $\varepsilon(u,i) = i$ and
\begin{align*}
  \X_a r(u,i) &= r(u, \min\set{j > i}{a_j = a}), \\
  \Y_{\!a} r(u,i) &= r(u, \max\set{j < i}{a_j = a}).
\end{align*}
If a nonempty ranker $r$ starts with an~$\X_a$-modality, then we say
that $r$ is an~\emph{$\X$-ranker}; and we define $r(u) = r(u,0)$,
\textit{i.e.}, the evaluation of $\X$-rankers starts at the beginning
of the word~$u$.  Symmetrically, if $r$ starts with $\Y_a$, then $r(u)
= r(u,n+1)$. As usual, $\min\emptyset$ and $\max\emptyset$ are
undefined. Thus a nonempty ranker $r$ either defines a unique position
$r(u)$ in a word $u$ or $r(u)$ is undefined. For example, $\X_a\!
\Y_{\hspace*{-0.2mm}b} \X_c (bac) = 3$ whereas $\X_a\!
\Y_{\hspace*{-0.2mm}b} \X_c (cba)$ is undefined.  For a ranker~$r$ we set
$L(r) = \set{u \in A^*}{\text{\ensuremath{r(u)} is defined}}$.
\ifjournalversion\alx{po2dfa-Charakterisierung von STV aus Thm unten
  mittels Technik von MK herleiten.}\fi
\begin{theorem}\label{thm:BGinDA}
  Let $L \subseteq A^*$. The following are equivalent:
  \begin{enumerate}
  \item\label{aaa:BGinDA} $L \in \calDA(A^*)$ is a
    Boolean combination of right ideals.
  \item\label{bbb:BGinDA} $L$ is a finite union of unambiguous
    monomials $A_1^* a_1 \cdots A_k^* a_k A_{k+1}^*$ \\ with
    $\smallset{a_i, \ldots, a_k} \nsubseteq A_i$ for all $i \in
    \smallset{1, \ldots, k}$.
  \item\label{ccc:BGinDA} $L$ is Boolean combination of
    languages $L(r)$ for $\X$-rankers~$r$.
  \item\label{ddd:BGinDA} $L$ is recognized by a
    one-pass po2dfa.
  \end{enumerate}
\end{theorem}
\begin{proof}
  Before turning to the actual proof, we give a rough overview of the
  techniques employed. %
  Right ideals are the finitary version of open sets in the Cantor
  topology over infinite words. It is therefore not surprising that a
  large part of Theorem~\ref{thm:BGinDA} reduces to infinite words:
  The proof of the implication from~\eqref{aaa:BGinDA}
  to~\eqref{bbb:BGinDA} relies on a result of Diekert and
  Kuf\-leitner~\cite[Theorem~6.6]{dk11tocs:short}. The step
  from~\eqref{bbb:BGinDA} to~\eqref{ccc:BGinDA} uses a
  characterization of $\X$-rankers over infinite
  words~\cite[Theorem~3]{dkl10dlt:short}. Showing the implication
  from~\eqref{ccc:BGinDA} to~\eqref{ddd:BGinDA} is the most technical
  part. In particular, one has to show that one-pass po2dfa are closed
  under union and intersection. Here, the respective result for
  po2-B\"uchi automata cannot be applied directly, but showing closure
  under union and intersection resembles techniques which were
  developed for deterministic po2-B\"uchi
  automata~\cite{kl11ijfcs:short}. This is
  Lemma~\ref{lem:det-po2-BooAlg}. Finally, the step
  from~\eqref{ddd:BGinDA} back to~\eqref{aaa:BGinDA} easily follows by
  combining the characterization of po2dfa due to Schwentick,
  Th{\'e}rien, and Vollmer~\cite[Theorem~3.1]{stv01dlt:short} with
  Theorem~\ref{thm:boolcantor}.  We need to introduce some more
  notation for the proof.

  A monomial $A_1^* a_1 \cdots A_k^* a_k A_{k+1}^*$ is
  \emph{restricted} if there exists no $i \in \smallset{1, \ldots, k}$
  such that $\smallset{a_i, \ldots, a_k} \subseteq A_i$. Let $\DA$ be
  the class of finite monoids satisfying the identity $(xy)^\omega =
  (xy)^\omega x (xy)^\omega$. A language $L$ is contained in $\calDA$
  if and only if it is recognized by a homomorphism $h: A^* \to M$ to
  a finite monoid in $\DA$.
  The set of finite and infinite words over $A$ is $A^\infty$. The
  $\omega$-iteration of a language $L \subseteq A^*$ of finite words
  is $L^\omega = \set{u_1 u_2 \cdots \in A^\infty}{u_i \in L}$; in
  particular we stipulate the convention $\varepsilon^\omega =
  \varepsilon$. Note that it will always be clear from the context
  whether by ``$\omega$'' we mean an infinite product or a generated
  idempotent.
  Let $h: A^* \to M$ be a homomorphism to a finite monoid $M$. For $x
  \in M$ let $[x] = h^{-1}(x)$.  A language $K \subseteq A^\infty$ of
  finite and infinite words is recognized by $h$ if 
  \begin{equation*}
    K = \bigcup \set{[s][e]^\omega}{[s][e]^\omega \isect L \neq
      \emptyset \text{ and }  s=se,\ e^2 = e}.
  \end{equation*}
  Note that $[1]^\omega$ also contains finite words.
  The evaluation of an $\X$-ranker~$r$ extends naturally to infinite
  words and the $L(r)$ over $A^\infty$ consists of all finite or
  infinite words on which $r$ is defined. For more details we refer
  to~\cite{dkl10dlt:short}.
  
  \eqref{aaa:BGinDA}\;$\implies$\;\eqref{bbb:BGinDA}: %
  By Theorem~\ref{thm:boolcantor} the language $L \subseteq A^*$
  is recognized by some $h: A^* \to M \in \DA$ such that $h(L)$ is a
  union of $\greenR$-classes. Consider the language
  \begin{equation*}
    K = \bigunion\set{[s][e]^\omega}{[s] \subseteq L \text{ and } s=se,\
    e^2 = e}
  \end{equation*}
  of finite and infinite words.  We have $K \isect A^* = L$ and~$K$ is
  recognized by~$h$. Consider $s,t,e,f \in M$ such that $s=se$,
  $t=tf$, $e^2 = e$ and $f^2 = f$.  Then because~$h(L)$ is a union of
  $\greenR$-classes, $s \Req t$ implies $[s][e]^\omega \subseteq K$ if
  and only if $[t][f]^\omega \subseteq K$.  The language $K$ is a
  finite union of restricted unambiguous monomials $A_1^* a_1 \cdots
  A_k^* a_k A_{k+1}^\infty$ over $A^\infty$,
  see~\cite[Theorem~6.6]{dk11tocs:short}. Therefore, $L = K \isect
  A^*$ is a finite union of restricted unambiguous monomials $A_1^*
  a_1 \cdots A_k^* a_k A_{k+1}^*$ over $A^*$.
  
  \eqref{bbb:BGinDA}\;$\implies$\;\eqref{ccc:BGinDA}: %
  Let~$L$ be a finite union of restricted unambiguous monomials of the
  form $A_1^* a_1 \cdots A_k^* a_k A_{k+1}^*$. Let~$K \subseteq
  A^\infty$ be obtained by replacing these monomials by the monomial
  $A_1^* a_1 \cdots A_k^* a_k A_{k+1}^\infty$. Then~$K$ is a union of
  restricted unambiguous monomials over $A^\infty$. Now, $K$ is
  definable over $A^\infty$ in the first-order fragment
  $\Delta_2[{<}]$, see~\cite[Theorem~6.6]{dk11tocs:short}. Thus $K$ is
  a Boolean combination of languages $L(r)$ over $A^\infty$ for
  $\X$-rankers $r$, see~\cite[Theorem~3]{dkl10dlt:short}. It follows
  that $L = K \isect A^*$ is a Boolean combination of languages $L(r)$
  over $A^*$ for $\X$-rankers~$r$.
  
  \eqref{ccc:BGinDA}\;$\implies$\;\eqref{ddd:BGinDA}: %
  It is easy to see that every language $L(r)$ for an $\X$-ranker~$r$
  is recognizable by a one-pass po2dfa. With
  Lemma~\ref{lem:det-po2-BooAlg}, we get closure of one-pass po2dfa
  under Boolean operations.
  
  \eqref{ddd:BGinDA}\;$\implies$\;\eqref{aaa:BGinDA}: %
  Let~$L$ be recognized by a complete one-pass po2dfa~$\calA$. In
  particular, $\calA$ is a po2dfa and thus
  $L \in \calDA$, \textit{cf.}~\cite[Theorem~3.1]{stv01dlt:short}.
  Let~$n$ be a number greater than the number of states of~$\calA$ and
  let $x, y, z \in A^*$. We claim that $z(xy)^n \in L$ if and only if
  $z(xy)^n x \in L$: Consider the run of~$\calA$ on either
  word. Let~$q$ be the state in which~$\calA$ leaves the prefix
  $z(xy)^n$ for the first time. Note that this must happen eventually
  since~$\calA$ is complete and the left end marker~$\leow$ cannot be
  trespassed. Then~$q$ is right-moving and $q \trans{a} q$ is a loop
  for all letters in $a \in \alph(xy)$ by choice of $n$. Hence,
  $\calA$ encounters the right end marker~$\reow$ in the state~$q$ on
  both inputs $z(xy)^n$ and $z(xy)^n x$. Therefore, $z(xy)^n$ is
  accepted if and only if $z(xy)^n x$ is accepted. By
  Theorem~\ref{thm:boolcantor}, the language $L$ is a Boolean
  combination of right ideals.
  \qed
\end{proof}

It is decidable whether a given regular language belongs to
$\calDA$. Therefore, using Proposition~\ref{prp:cantor} and
Theorem~\ref{thm:boolcantor}, it is decidable whether a regular
language is recognized by an arbitrary (resp.\ flip, fully final)
one-pass po2dfa.
The temporal logic version of $\X$-rankers is denoted
$\TL_{\X}[\X_a,\Y_{\!a}]$, \textit{cf.}~\cite{dkl10dlt:short}; it is a
fragment of deterministic unary temporal logic $\TL[\X_a,\Y_{\!a}]$
over the modalities~$\X_a$ and~$\Y_{\!a}$. The logic
$\TL[\X_a,\Y_{\!a}]$ is expressively complete for $\calDA$, and
$\TL_{\X}[\X_a,\Y_{\!a}]$ defines the right ideals in $\calDA$.

\begin{remark}\label{rem:po2hierarchy}
  We use the shortcut ``$\mathrm{nfa}$'' for \emph{nondeterministic
    finite automaton}, and ``$\mathrm{po1}$'' for \emph{partially
    ordered one-way}. Using this notation, we have the following
  inclusions between language classes recognizable by partially
  ordered automata:
  \begin{align*}
    \mathrm{po1dfa} \ \subsetneq \ 
    \mathrm{one\mbox{-}pass} \ \mathrm{po2dfa} \ \subsetneq \
    \mathrm{po2dfa} \ \subsetneq \
    \mathrm{po2nfa} \ = \
    \mathrm{po1nfa}.
  \end{align*}
  The following (very similar) languages show that the inclusions are
  strict.  The language $\oneset{a,c}^* ab \oneset{a,b,c}^*$ is
  recognizable by some one-pass $\mathrm{po2dfa}$ but not by a
  $\mathrm{po1dfa}$. The language $\oneset{a,b,c}^* ab \oneset{b,c}^*$
  is recognizable by a $\mathrm{po2dfa}$ but not by any one-pass
  $\mathrm{po2dfa}$. Finally, the language $\oneset{a,b,c}^* ab
  \oneset{a,b,c}^*$ is recognizable by some $\mathrm{po1nfa}$ but not
  by any $\mathrm{po2dfa}$. The equivalence of $\mathrm{po2nfa}$ and
  $\mathrm{po1nfa}$ is due to Schwentick, Th{\'e}rien, and
  Vollmer~\cite{stv01dlt:short}.  For each of the above language
  classes the membership problem is decidable: The class
  $\mathrm{po1dfa}$ corresponds to $\greenR$-trivial
  monoids~\cite{stv01dlt:short}, one-pass $\mathrm{po2dfa}$ correspond
  to $\greenR$-classes of monoids in~$\DA$
  (Theorem~\ref{thm:boolcantor} and Theorem~\ref{thm:BGinDA}). The
  algebraic equivalent of $\mathrm{po2dfa}$ is the variety of finite
  monoids $\DA$~\cite{stv01dlt:short}, and $\mathrm{po2nfa}$ are
  expressively complete for the level 3/2 of the Straubing-Th{\'e}rien
  hierarchy~\cite{stv01dlt:short} which is decidable by a result of
  Pin and Weil~\cite{pw97:short}.
 \eex
\end{remark}

In analogy to Theorem~\ref{thm:BGinDA}, there is also an
expressively complete two-way automaton model for Boolean combinations
of right ideals. A two-way automaton is \emph{weak} if for every strongly
connected component either all states are final or all states are
non-final. 
Note that every partially ordered automaton is weak. The following
result is our only general result for arbitrary (not
partially ordered) deterministic two-way automata.
\begin{proposition}\label{prp:weak2way}
  A regular language is a Boolean combination of right ideals if and
  only if it is recognized by a deterministic weak one-pass two-way
  automaton.
  \ifjournalversion\alx{Example of a \emph{non}deterministic weak one-pass
    two-way automata not recognizing a Boolean combination of right
    ideals?}\fi
\end{proposition}
\begin{proof}
  If $L$ is a Boolean combination of right ideals, then $L$ is
  recognized by a deterministic weak (one-way) automaton by
  Theorem~\ref{thm:weakdfa}. Note that every one-way automaton can
  also be seen as a two-way automaton without left-moving states.
  
  For the converse,  consider a complete deterministic weak
  one-pass two-way automaton~$\calA$. By
  Theorem~\ref{thm:boolcantor} it suffices to show that $L(\calA)$
  satisfies the lattice identity $z(xy)^\omega \,\liff\,
  z(xy)^\omega x$. 
  The \emph{leaving state} of $u$ is the state of~$\calA$ which on
  input~$u$ encounters the right end marker~$\reow$ for the first
  time. Note that, since $\calA$ is complete and deterministic, there
  is a unique state with this property.
  Consider words $x, y, z \in A^*$. There are only finitely many
  strongly connected components of~$\calA$. Consequently the
  pigeonhole principle yields an integer~$n$ such that the leaving
  states of $z (xy)^n$ and of $z (xy)^{n+1}$ are in the same strongly
  connected component. Hence, the same is true for the leaving states
  $p$ of $z (xy)^n$ and $q$ of $z (xy)^n x$. Since $\calA$ is weak,
  $p$ is final if and only if $q$ is; and because $\calA$ is a
  one-pass automaton we have $z (xy)^n \in L(\calA)$ if and only if $z
  (xy)^n x \in L(\calA)$. This establishes the lattice identity.
  \qed
\end{proof}

Not every deterministic one-pass two-way automaton recognizing a
Boolean combination of right ideals needs to be weak. Therefore, the
equivalence of \eqref{aaa:BGinDA} and \eqref{ddd:BGinDA} in
Theorem~\ref{thm:BGinDA} does not follow from
Proposition~\ref{prp:weak2way}.  Also note that the analogue of
Proposition~\ref{prp:weak2way} does not work for right ideals (resp.\
prefix-closed languages) and deterministic flip (resp.\ fully accepting)
one-pass two-way automata since deterministic two-way automata can
also reject an input by an infinite cycle in its computation.

\ifjournalversion\alx{Ist dies das einzige Problem? \mkk{Wahrscheinlich ja;
    aber da hab' ich jetzt nicht die Zeit und die Nerven 'fuer.} Wenn
  ja, dann koennte es helfen Automaten zu betrachten, so dass fuer
  jede SCC alle Zustaende entweder $X$ oder $Y$ sind (aber nicht
  beides vorkommt). (Dies liefert wohl auch eine natuerliche
  Zwischenklasse zwischen ``weak one-way'' und ``weak two-way''.)
  \mkk{Ist schwer abzuschaetzen, ob das interessant ist; solche
    Automaten koennen nur beschraenkt oft die Richtung wechseln (und
    fuer groessere Automatenklassen koennte es sich evtl dadurch zu
    leicht auf den one-way Fall reduzieren).}} \alx{Guter Punkt. Es
  ist auf jeden Fall etwas, worueber man bei einer etwaigen
  Zeitschriftenversion nachdenken kann.}\fi

As for one-way automata in Section~\ref{sec:automata}, we get right
ideals in $\calDA$ if the recognizing automaton is a flip
automaton. For a \emph{flip automaton}, a transition $z \trans{a} z'$
with final state $z$ implies that $z'$ is final. As an intermediate
step, we get a characterization in terms of unambiguous monomials.
\begin{theorem}\label{thm:GinDA}
  Let $L \subseteq A^*$. The following are equivalent:
  \begin{enumerate}
  \item\label{aaa:GinDA} $L \in \calDA(A^*)$ is a right ideal.
  \item\label{bbb:GinDA} $L$ is a finite union of unambiguous
    monomials $A_1^* a_1 \cdots A_k^* a_k A^*$.
  \item\label{ddd:GinDA} $L$ is recognized by a complete flip one-pass
    po2dfa.
 \end{enumerate}
\end{theorem}
\begin{proof}
  We first show \eqref{aaa:GinDA}\;$\implies$\;\eqref{ddd:GinDA}.
  Suppose $L \in \calDA(A^*)$ is a right ideal of~$A^*$. By
  Theorem~\ref{thm:BGinDA} there exists a complete one-pass
  po2dfa~$\calA$ which recognizes~$L$. We show how to obtain an
  equivalent automaton~$\calB$ which is a flip automaton.
  Let us say that, during a computation, a deterministic automaton is
  in \emph{progress mode} if after the next transition is taken, the
  automaton scans a position which has not been scanned before. The
  idea is that we need to change into a final state only when in
  progress mode.  Note that for acceptance, the crucial transition of
  a (one-pass) two-way automaton is always made in progress
  mode. Moreover, consider an input $uav$ and suppose~$\calA$ scans
  position $\abs{ua}$ in progress mode and performs a transition into
  a final state. Then the prefix $ua$ is accepted because~$\calA$ is a
  one-pass automaton (note that this would not hold if $\calA$ has
  already seen some prefix of $v$ during the computation).  Since $L$
  is a right ideal, all words in $uaA^*$ are accepted. This shows that
  if in progress mode a transition into a final state is made, then we
  can directly go into a final, right-moving sink state without
  changing the language. In total this yields a complete one-pass
  po2dfa which is a flip automaton. It remains to show that we can
  simulate~$\calA$ in such a way that the simulation is aware of when
  it is in progress mode.

  Assume that~$\calA$ is leaving progress mode. This can only happen
  by a transition to a left-moving state. Suppose the input is
  factorized as $u = u_1 a_1 \cdots u_m a_m u'$ where the $a_i$'s
  correspond to the positions where a state change happened while in
  progress mode. Note that $a_m$ corresponds to the position scanned
  before taking the transition because a state change is necessary to
  leave progress mode. Now since~$\calA$ is deterministic, we see $a_i
  \nin \alp(u_i)$ for all $i$. Moreover since~$\calA$ is partially
  ordered, $m$ is bounded by the number of states
  of~$\calA$. Therefore, the simulation can store the word $v = a_1
  \cdots a_m$ in a stack of letters with bounded depth in its state
  space.
  Using the automaton $\calC$ of Lemma~\ref{lem:autopf} with $v = a_1
  \cdots a_m$, we can simulate~$\calA$ in the subsequent
  non-progressing phase and recognize when we are scanning the
  frontier $a_m$ of progress again.  The automaton is complete and
  thus there eventually is a transition trespassing the position
  corresponding to~$a_m$. This is when the simulation switches back to
  progress mode. Back in progress mode, the simulation organizes the
  stack by pushing the currently scanned letter to the stack if it
  causes a state change.

  \eqref{ddd:GinDA}\;$\implies$\;\eqref{bbb:GinDA}: %
  Let $L = L(\calA)$ for a complete one-pass po2dfa which is a flip
  automaton.  For every $u \in L(\calA)$ we construct an unambiguous
  monomial $P(u) = A_1^* a_1 \cdots A_k^* a_k A^*$ such that $u \in
  P(u) \subseteq L(\calA)$ and~$k$ bounded by the number of states
  of~$\calA$. Since there are only finitely many such monomials, we
  have $L(\calA) = \bigunion_{u \in L} P(u)$ and this union is finite.

  To construct $P(u)$ consider $u \in L(\calA)$ and fix an accepting
  computation of~$\calA$ on~$u$. Consider the factorization $u = u_1
  a_1 \cdots u_k a_k u'$ where the $a_i$'s correspond to state
  changes. Let $P(u) = A_1^* a_1 \cdots A_k^* a_k A^*$ with $A_i =
  \alp(u_i)$. Trivially, $u \in P(u)$ and $k$ is bounded by the number
  of states of~$\calA$. Moreover, $P(u)$ is unambiguous because
  $\calA$ is deterministic.
  It remains to show $P(u) \subseteq L(\calA)$. Suppose~$\calA$ is in
  some state $z$ while scanning a $b$-position of $u_i$. By
  construction there is no state change with the next transition,
  \textit{i.e.}, there is a loop $z \trans{b} z$ in~$\calA$.
  Consider some word $v \in P(u)$ and factorize $v = v_1 a_1 \cdots
  v_k a_k v'$ with $v_i \in A_i^*$. By construction, there exists a
  run of~$\calA$ on $v$ which eventually trespasses $a_k$ into a final
  right-moving state. Then since~$\calA$ is a complete flip automaton,
  no matter what comes beyond the $a_k$ can remedy acceptance. This
  shows that every $v \in P(u)$ is accepted.

  \eqref{bbb:GinDA}\;$\implies$\;\eqref{aaa:GinDA}: Every union of
  unambiguous monomials is in $\calDA$,
  \textit{cf.}~\cite{tt02:short,dgk08ijfcs:short}. By
  Proposition~\ref{prp:cantor}, we see that $L$ is a right ideal.
%
  \qed
\end{proof}

Note that property~\eqref{bbb:GinDA} in Theorem~\ref{thm:GinDA} states
that unambiguity of monomials and the ideal property can be achieved
simultaneously, which is non-trivial.
A two-way automaton is \emph{fully accepting} if all its states are
final. As for one-way automata, this yields prefix-closed languages
(at least for $\calDA$).  The following result for prefix-closed
languages is an immediate corollary of Theorem~\ref{thm:GinDA}.

\begin{corollary}\label{cor:FinDA}
  Let $L \subseteq A^*$. The following are equivalent:
  \begin{enumerate}
  \item\label{aaa:FinDA} $L \in \calDA(A^*)$ is prefix-closed.
  \item\label{ddd:FinDA} $L$ is recognized by a fully accepting
    one-pass po2dfa.
    \qed
  \end{enumerate}
  \ifjournalversion\alx{UPols for filters in DA?}\fi
\end{corollary}
\begin{proof}
  \eqref{aaa:FinDA}\;$\implies$\;\eqref{ddd:FinDA}: %
  Let $L \in \calDA(A^*)$ be prefix-closed. The complement $A^*
  \setminus L$ is a right ideal in $\calDA(A^*)$ and thus
  Theorem~\ref{thm:GinDA} yields a complete one-pass po2dfa $\calA =
  (Z, A, \delta, x_0, F)$ which is flip and recognizes $A^* \setminus
  L$. We can assume $x_0 \not\in F$ since otherwise $\varepsilon \in
  A^* \setminus L$ and thus $A^* \setminus L = A^*$ and $L =
  \emptyset$ (and for $L = \emptyset$ we allow the empty automaton).
  Let $\calA' = (Z \setminus F, A, \delta', x_0, Z \setminus F)$ be
  the deterministic one-pass \potwo-automaton obtained from~$\calA$ by
  restricting the states to $Z \setminus F$, \textit{i.e.}, the
  transition relation $\delta'$ is given by $z \trans{a} z'$ in
  $\calA'$ if $z, z' \in Z \setminus F$ and $z \trans{a} z'$
  in~$\calA$. Clearly, $\calA'$ is fully accepting and a
  straightforward verification yields $L(\calA') = A^* \setminus
  L(\calA)$.
  
  \eqref{ddd:FinDA}\;$\implies$\;\eqref{aaa:FinDA}: %
  Suppose $L = L(\calA)$ for a fully accepting one-pass po2dfa $\calA
  = (Z, A, \delta, x_0, Z)$. Let $\calA' = (Z \mathbin{\dot\union}
  \oneset{x_f}, A, \delta', x_0, \oneset{x_f})$ where $x_f$ is a new
  right-moving sink state, \textit{i.e.}, $\delta'$ extents $\delta$
  with transitions $z \trans{a} x_f$ for $z \in Z \union \oneset{x_f}$
  if there exists no $z' \in Z$ such that $(z, a, z') \in
  \delta$. Then $\calA'$ is a complete one-pass flip po2dfa and
  $L(\calA')$ is a right ideal in~$\calDA(A^*)$ by
  Theorem~\ref{thm:GinDA}. Since $L(\calA) = A^* \setminus L(\calA')$,
  we see that $L \in \calDA(A^*)$ is prefix-closed.
  \qed
\end{proof}

\paragraph{\textbf{Acknowledgments.}}
We thank the anonymous referees for several suggestions which helped
to improve the presentation of the paper, and we are also grateful for
bringing to our attention the works of Avgustinovich and
Frid~\cite{AvgustinovichFrid06csr:short} and of Paz and
Peleg~\cite{PazPeleg65jacm:short}.


\newcommand{\Ju}{Ju}\newcommand{\Ph}{Ph}\newcommand{\Th}{Th}\newcommand{\Ch}{Ch}\newcommand{\Yu}{Yu}\newcommand{\Zh}{Zh}

\end{document}